\documentclass[11pt,twoside]{article}
\usepackage{asp2010}

\resetcounters

\usepackage{url}

\def\arcs{\hbox{$^{\prime\prime}$}}        
\def\farcs{\hbox{$.\!\!^{\prime\prime}$}}  
\def\mm{\hbox{$^{\rm m}$}}                 
\def\fmm{\hbox{$.\!\!^{\rm m}$}}           

\begin{document}

\title{The New Milky Way: 
a wide--field survey of optical transients\\ near the Galactic plane}
\author{Kirill~Sokolovsky,$^{1,2}$ Stanislav~Korotkiy,$^3$ and Alexandr~Lebedev$^4$
\affil{$^1$Astro Space Center of Lebedev Physical Institute, Russian Academy
of Sciences, Profsoyuznaya~Str.~84/32, 117997 Moscow, Russia}
\affil{$^2$Sternberg Astronomical Institute, Moscow State University,
Universitetskii~pr.~13, 119992~Moscow, Russia}
\affil{$^3$``Ka-Dar'' astronomy foundation, Kuzminki, P.O.~Box~82, 142717~Moscow~region, Russia}
\affil{$^4$Institute of Astronomy, Russian Academy of Sciences,
Pyatnitskaya~Str.~48, 119017 Moscow, Russia}}

\begin{abstract}
Currently, it may take days for a bright nova outburst to be detected. 
With the few exceptions, little is known about novae behaviour prior to maximum light.
A theoretically-predicted population of ultra-fast novae with
$\mathrm{t}_2<1^\mathrm{d}$ 
is evading observational discovery because it is not possible to
routinely organize fast follow-up observations of nova candidates.
With the aim of brining the detection time of novae and other bright ($V<13.5$) 
optical transients from days down to hours or less, we develop an automated
wide-field ($8^\circ\times6^\circ$) system capable of surveying the whole Milky Way
area visible from the observing site in one night.
The system is built using low-cost mass-produced components and the
transient detection pipeline is based on the open source \texttt{VaST} software.
We describe the instrument design and report results of the first 
observations conducted in October--November 2011 and January--April 2012.
The results include the discovery of Nova~Sagittarii~2012~No.~1 as well as 
two X-ray~emitting cataclysmic variables 1RXS~J063214.8$+$25362 and 
XMMSL1~J014956.7$+$533504. The rapid detection of Nova~Sagittarii~2012~No.~1
enabled us to conduct its X-ray and UV observations with {\em Swift} 22~hours
after discovery ($\simeq 31$~hour after the outburst onset).
All images obtained during the transient search survey are available online.
\end{abstract}

\section{Introduction}

A transient detection survey is characterized 
by its depth, area and cadence. 
Most of the current optical surveys are either deep/low-cadence 
\citep{2007ARep...51.1004L,2009ApJ...696..870D,2010NewA...15..433M,2010SPIE.7733E..12K,2013NewA...19...99H} 
or very shallow/very high-cadence 
\citep{2005NewA...10..409B,2010AstBu..65..286B,2010fym..confP..55U}. 
The gap between the two extremes (a medium-depth/medium-cadence survey) is covered by numerous
transiting exoplanets search experiments 
\citep[e.g.,][]{2003ASPC..294..405S,2004ApJ...613L.153A,2005PASP..117..783M,2007PASP..119..923P,2009IAUS..253..354B} 
which typically lack real-time transient detection capability and amateur
nova searches. 
Two notable examples of wide-field medium-depth surveys capable of detecting transients
are ASAS-3 
\citep{2001ASPC..246...53P} and
ROTSE-I/NSVS 
\citep{1998APS..APR..R802L,2004AJ....127.2436W}, unfortunately ROTSE-I has
completed its observations in 2001 and ASAS-3 
has not reported detection of a transient source since 2011. 
There are, however, good reasons to pay more attention to transient 
search in this region of survey parameter space.

Potential targets for a medium-cadence (about one visit per night) optical survey in the 9--14$^\mathrm{m}$ range
include classical and symbiotic novae, cataclysmic variables (dwarf
novae) and other X-ray binaries, FU~Ori-type flares, brightest microlensing events
\citep{2006ATel..943....1M},
intriguing ``anti-transients'' characterized by a dramatic drop in star's
brightness not related to an eclipse or R~CrB-type dust formation event
\citep{2012ATel.4784....1D},
other variable stars, bright blazar flares. Rapid detection of extreme
events such as star-star \citep{2006A&A...451..223T,2011A&A...528A.114T} and star-planet
\citep{2012MNRAS.425.2778M} mergers or a Galactic supernova
\citep{2011ApJ...737L..22C} would be of crucial importance for their investigation.

The expected classical novae rate in the Galaxy is $\sim 35$~yr$^{-1}$
\citep{1997ApJ...487..226S,2006MNRAS.369..257D}, but the actual discovery
rate\footnote{\url{http://www.cbat.eps.harvard.edu/nova_list.html}} is 
$\sim 10$~yr$^{-1}$.
While some are missed due to their Sun proximity, a number
of potentially detectable outbursts are missed because current searches lack time
coverage and depth.
If a typical outburst detection time could be shortened from days down to a few hours or
less, it would provide obvious scientific benefits.
Nova detection before its maximum light ensures accurate
determination of its maximum magnitude and decline rate ($t_2$, $t_3$) 
and the values derived from them.
Better constrained outburst onset time would aid interpretation of unusual
cases such as $\gamma$-ray-bright novae
\citep{2010Sci...329..817A,2012ATel.4284....1C,2012ATel.4310....1C}. 
Rise time distribution of novae
could be studied if accurate outburst onset times are known for many of them.
A theoretically-predicted population of ultra-fast novae with
$\mathrm{t}_2<1^\mathrm{d}$ is evading detection because 
it is not possible to routinely organize fast follow-up observations 
of nova candidates.

We describe the ``New Milky Way'' (NMW) -- an automated system capable of 
surveying the whole Milky Way area visible from the observing site during one night
down to magnitude $V<13.5$. The system is built using low-cost components 
and the transient detection pipeline is based on open source software.
The survey strategy and data reduction pipeline are designed with the aim to 
inform the community and enable multi-wavelength follow-up of detected transients 
a few hours after they are imaged.

\section{The New Milky Way camera}

The transient detection system consists of a wide-field survey camera:
an f$=135$~mm f/2.0 telephoto lens attached to an unfiltered ST8300M CCD
installed on a HEQ-5~Pro robotic mount, a \texttt{Windows}-based computer controlling
the mount and the camera through a custom-made software developed by 
Vasily Vershinin and a \texttt{Linux}-based data-reduction computer.
The system is operated from an altitude of 2000~m by Ka-Dar Observatory at Karachay-Cherkessia 
(MPC code C32) at the Russia's North Caucasus mountains. The system parameters are
summarized in Table~\ref{tab:system}.

\begin{table}[t!]
\setlength{\tabcolsep}{4pt}
\begin{center}
\caption{The New Milky Way survey parameters.}
\label{tab:system}
\smallskip
{\small
\begin{tabular}{lr|lr}
  \hline                        
  \hline
  Optics:             & Canon 135~mm f/2.0 &           Accuracy at $V\sim11$: & $\sim0\fmm1$ absolute \\
  CCD camera:         & SBIG ST-8300M &                                       & $0\fmm05$  internal \\
                      & (blue-sensitive chip) &        Images per field: & 2 or 3 per night \\
  Image size:         & 3352$\times$2532 pix &                                     & (dithering) \\
  Optical filter:     & none &                         Stars per frame: & $\sim20000$ \\
  Equatorial mount:   & HEQ-5~Pro &                    Images per night: & $\sim200$ \\
  Field of view:      & $8^\circ \times 6^\circ$ &     Milky Way imaging time: & 5~hr (January) \\
  Pixel scale:        & 8.4$\arcsec$/pix &                                           & 10~hr (April) \\
  Exposure time:      & 20--40~sec &                   Processing time: & up to 7~hr/night \\
  Limiting magnitude:  & $V\sim14.5$ &                 Results inspection time: & up to 4~hr/night \\
  Transient detection limit: & $V\sim13.5$ & & \\
  \hline  
\end{tabular}}
\end{center}

\end{table}

\begin{figure}[]
    \includegraphics[height=4.75cm]{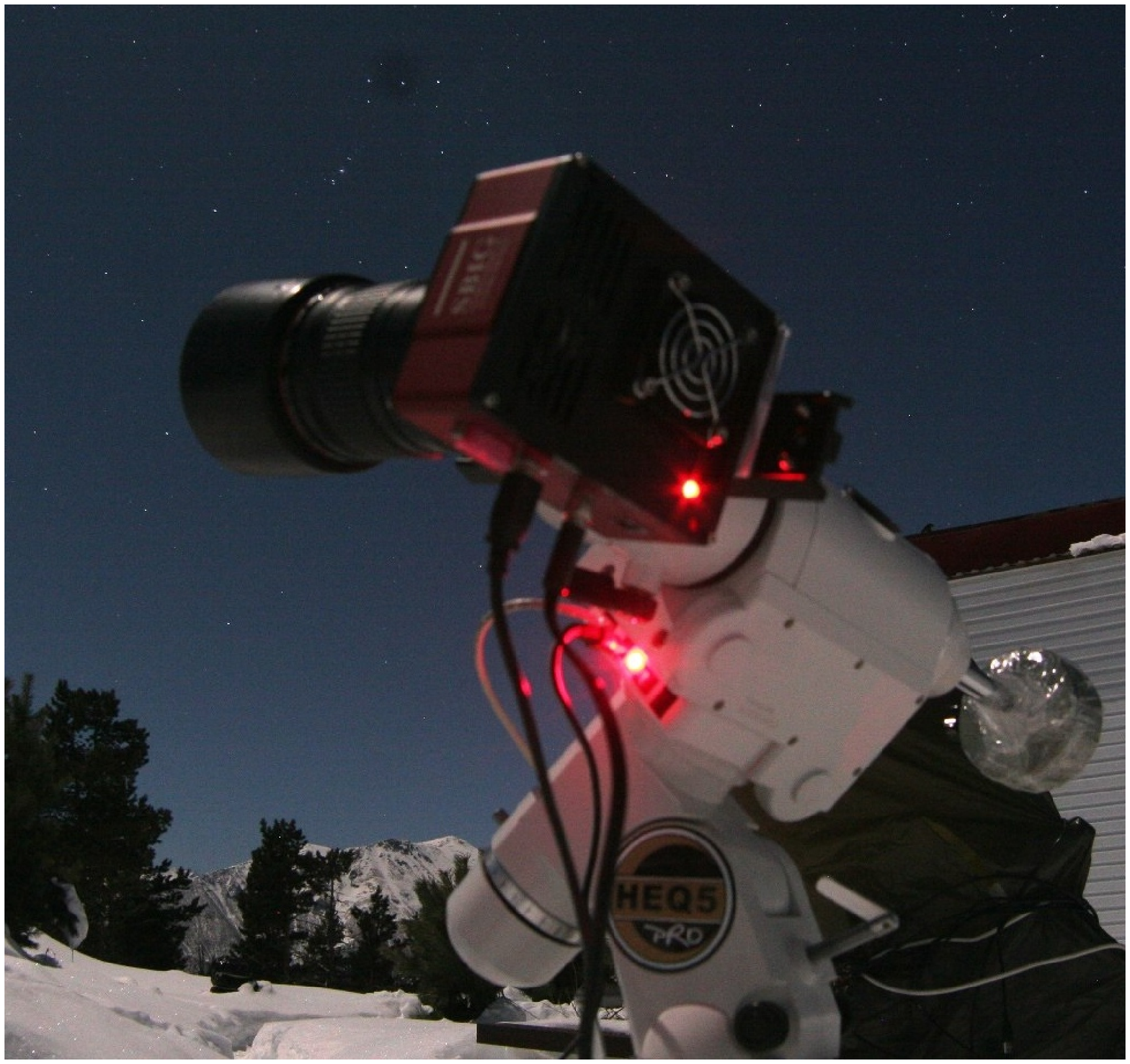}
    \includegraphics[height=4.75cm]{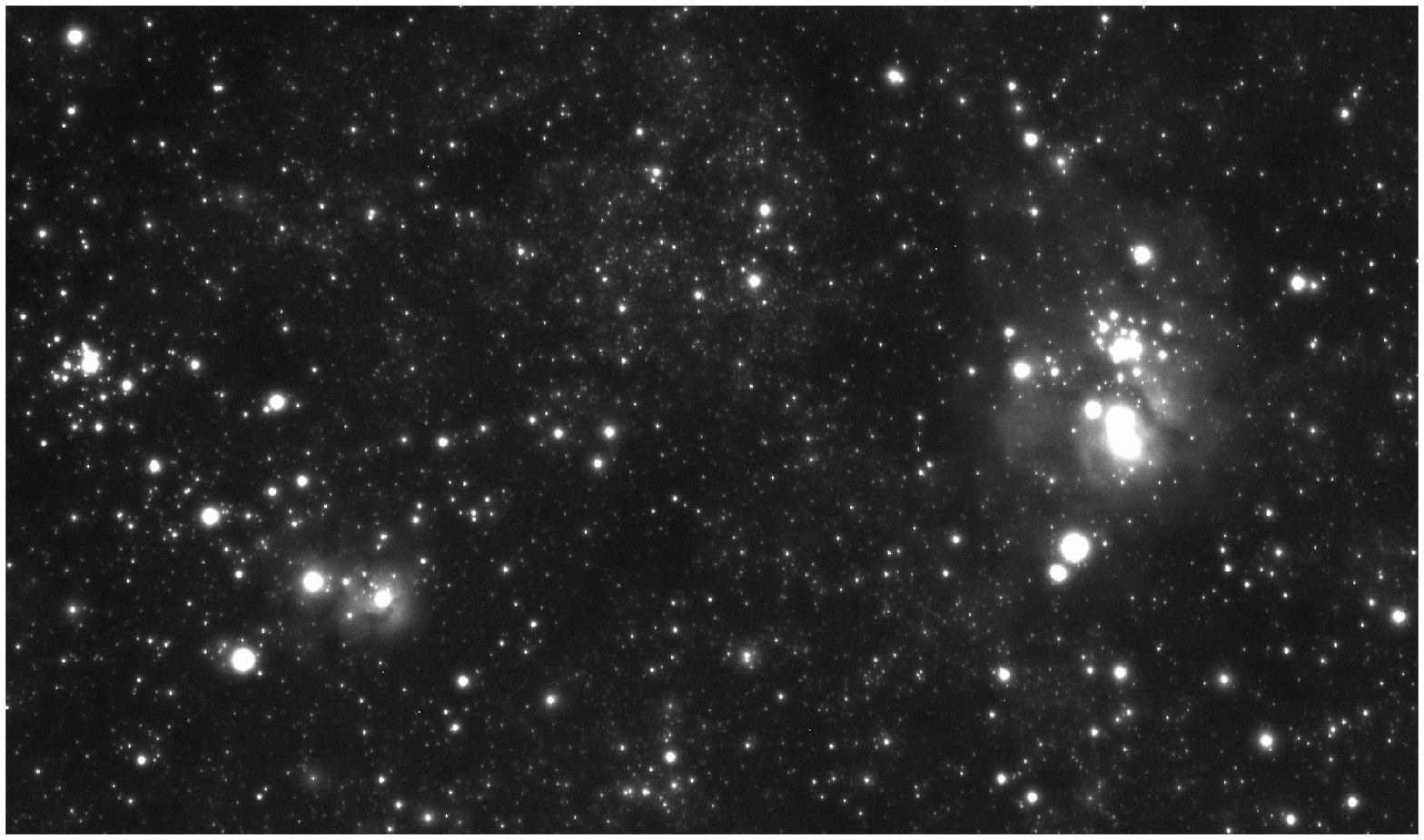}
  \caption{The New Milky Way camera (left) and an example image (right; fraction of the full
frame) showing the Lagoon (M8) and Trifid (M20) nebulae region in Sagittarius.}
  \label{fig:camera}
\end{figure}

For data reduction we choose object detection rather than
image subtraction, because the former approach is computationally more efficient and
provides better photometric accuracy despite being less sensitive in crowded
fields \citep{2012IAUS..285....9S}.
The transient detection pipeline is implemented as a \texttt{BASH} script controlling the
\texttt{VaST} software\footnote{\texttt{VaST} is the open source software available at \url{http://scan.sai.msu.ru/vast/}}
\citep{2005ysc..conf...79S}
which performs the following steps. {\it (i)}~Star detection and circular aperture photometry
using \texttt{SExtractor} \citep{1996A&AS..117..393B}. {\it (ii)}~Cross-matching the
lists of stars detected on reference (first-epoch) and second-epoch images. Transient
candidates are identified as objects detected on all second-epoch images
that were either not detected or were at least 1$\mm$ fainter on the reference image. 
The second criterion is needed to identify a low-amplitude flare
or a high-amplitude flare of a faint object that is blended with a
brighter one visible on the reference image. 
Small movement of the telescope between images (dithering)
is used to reject CCD artifacts. {\it (iii)}~Images are plate-solved using the   
\texttt{Astrometry.net}\footnote{see \url{http://astrometry.net/}} software
\citep{2010AJ....139.1782L,2008ASPC..394...27H} and celestial positions of
the detected stars obtained. {\it (iv)}~The instrumental magnitude scale is calibrated using V
magnitudes of unsaturated Tycho-2 \citep{2000A&A...355L..27H} stars in the field
of view.

The main problem of our transient detection system is that it currently
lacks an automated enclosure. While being able to perform
observations automatically, it still has to be set up, started and stopped manually. 
Thus, observing time is limited by the availability of a
qualified observer on site.

\section{Results}


{\bf Nova~Sagittarii~2012~No.~1.} Nova~Sgr~2012~No.~1 (PNV J17452791$-$2305213, \\ 
17:45:28.02 $-$23:05:23.1
$\pm0\farcs1$, J2000, \citealt{2012CBET.3089....1K}) was detected on the NMW images obtained on 2012
April~21.0112~UT at $V=9.6$ and reported through the CBAT Transient Objects
Confirmation Page
(TOCP\footnote{\url{http://www.cbat.eps.harvard.edu/unconf/followups/J17452791-2305213.html}}).
First independent confirmation images were obtained on April 21.654 UT by
J.~Seach (Australia) with a 50~mm f/1.0 lens attached to a DSLR
camera. The pre-discovery images obtained on April~20.84032~UT
at Xingming observatory (near Urumqi, China; MPC code C42) showing the
object at $V=10.2$ were later reported. It was discovered by A.~Watson that the outburst
was recorded in detail by the inner Heliospheric Imager telescope on the
STEREO~{\it Behind}
spacecraft\footnote{\url{http://stereo.gsfc.nasa.gov/~thompson/nova_sagittarii_2012/}}.
The STEREO-B observations allow to estimate the outburst onset time as April~$\sim20.6$~UT.
Availability of pre-discovery images by Xingming obs. and STEREO-B obtained
hours before the NMW images highlight the challenges of rapid information extraction 
from the data.

A {\it Swift} ToO observation was performed on
April~21.912, 22~hours after discovery and $\simeq 31$~hour after the outburst
onset \citep{2012ATel.4061....1S}.
To the best of our knowledge, this is the fastest X-ray follow-up
observation of an optically-detected nova. While the first and the two
following (April~25.7 and 26.1, \citealt{2012ATel.4088....1N}) observations
detected no X-ray emission, observations on May~10 revealed hard X-ray
emission from the nova shell \citep{2012ATel.4110....1N}. 
Optical spectroscopy  allows one to classify the
object as a fast He/N type nova with expansion velocity up to 6500~km/s 
\citep{2012CBET.3089....1K,2012ATel.4094....1E}.
Nova~Sgr~2012~No.~1 was also the target of radio
\citep{2012ATel.4088....1N} and infrared
observations \citep{2012ATel.4093....1R,2012ATel.4142....1V}.


{\bf 1RXS~J063214.8$+$25362.} A previously undetected $13\fmm1$ source appeared at \\
06:32:13.082 $+$25:36:22.68 ($\pm0\farcs1$, J2000) on 2012 January~24 gradually brightening to $12\fmm6$ by 
January~27 \citep{2012ATel.3893....1K}. The object was below the detection limit on January~21.
It is located $24\arcs$ from an unidentified X-ray source 
1RXS J063214.8$+$253620 listed in the
ROSAT All Sky Survey Faint Source Catalog \citep{2000IAUC.7432R...1V}
and is likely associated with it. The ASAS-3 database
has two detections of a source at the above position on 2009 March 17 and
21 at $V=12.9$ suggesting the object is a dwarf nova. Follow-up
spectroscopy by N.~Masetti (private comm.) revealed a spectrum
characteristic of a non-magnetic cataclysmic variable. 
Time series photometry reported in
vsnet-alert\footnote{\url{http://ooruri.kusastro.kyoto-u.ac.jp/mailarchive/vsnet-alert}} \#14161, \#14164, \#14165, and \#14210
suggested a presence of $0\fd32$ unstable periodic modulation.


{\bf XMMSL1~J014956.7$+$533504.} A $V=12.8$ transient was detected on 2012 January 29 at 01:49:56.77 
$+$53:35:01.8 ($\pm0\farcs5$, J2000, \citealt{2012ATel.3896....1K}) coinciding with an X-ray source listed 
in the XMM-Newton slew survey catalog XMMSL1 J014956.7+533504
\citep{2008A&A...480..611S}. Spectroscopy by A.~Arai, M.~Nagashima, C.~Naka
(Kyoto Sangyo Univ.) reported through TOCP\footnote{\url{http://www.cbat.eps.harvard.edu/unconf/followups/J01495677+5335018.html}}
revealed that the flare is a dwarf nova outburst.


{\bf Image archive.} The images obtained within the NMW survey are available at the
project's webpage\footnote{\url{http://scan.sai.msu.ru/nmw/}}. The archive
server is written in \texttt{Haskell} functional programming language with 
\texttt{Yesod} web framework by one of the
authors (AL) and is using \texttt{SWarp} \citep{2002ASPC..281..228B} for fast
image resampling.
The archive of photometric measurements is currently under development.

\section{Summary}

We developed and tested the NMW system capable of detecting
optical transients near the Galactic plane down to $V<13.5$ hours after
they appear. The design is based on relatively cheap mass-produced
components and open source image analysis software (partly developed by us), 
so similar transient-search systems can be easily deployed elsewhere.
Construction of an automated clam-shell or dome-type
enclosure is required to put our system on constant alert. 
The first survey results include the discovery of Nova~Sgr~2012~No.~1
and two X-ray~emitting cataclysmic variables 1RXS~J063214.8$+$25362 and
XMMSL1~J014956.7$+$533504.
A web-accessible image and photometry archive is
being developed to ensure full use of collected data.

\acknowledgements The authors thank Sergei Antipin, Maria Mogilen and Olga
Sokolovskaya for critically reading this manuscript.

\bibliographystyle{asp2010}
\bibliography{new_milky_way}

\end{document}